# Raman spectra and infrared intensities of graphene-like clusters in compared to epitaxial graphene on SiC

Seyed Sajad Sadeghi[1*], Hamidreza Simchi[2]

[1]Department of Physics, Bu-AliSina University,Hamedan, Iran.

[2]Department of Physics, Iran University of Science and Technology, Narmak, Tehran 16844, Iran.

**Abstract**

There are several growing methods for graphene. In this study, the growth of graphene-like clusters on the SiC wafers is done by annealing the wafers in a vacuum evaporation system equipped with a heating source accessory. For evaluating the quality of the growth method, the Raman spectra and infrared intensities of graphene-like clusters are studied theoretically and experimentally. For doing the theoretical study, three types of graphene clusters are considered and their Raman spectrum and infrared intensities are found using the Hartree-Fock method. The results show that the geometry of the cluster, and in consequence the geometry-dependent high (low) non-uniformity of charge distribution on the cluster surfaces causes the high (low) infrared intensities. The experimental spectrums are measured and compared with the theoretical ones. An agreement was seen between the experimental and theoretical Raman spectrum when the wave number is less than 1700 $Cm^{-1}$. It is shown that more accurate temperature control and higher vacuum level of the chamber are essential for using the physical evaporation method for growing the single-layer graphene on the SiC substrate.

**PACS Numbers: 81.05.ue, 87.64.Je, 81.15.−z**

**Keywords: Graphene Clusters, Epitaxy Method, Hartree-Fock, Raman Spectrum, Infrared Intensity, SiC Wafer**

## 1. Introduction:

The interest of technologists in using graphene as the main material for Nano-electronic applications has increased dramatically during recent years [1, 2]. The high mobility, tunability of transport characteristic, superior mechanical strength, and high thermal conductivity [3–7] of graphene makes it attractive for applications in Nano-scale electronic and spintronic devices. It should be mentioned that the lack of an energy gap in graphene is one of the significant hurdles for graphene device applications [8]. There are several preparation methods for graphene. Mechanical exfoliation [9], reducing graphene oxide [10, 11] (GO), chemical vapor position (CVD) [12, 13], and epitaxial growth on SiC substrate [14, 15] are some examples. The advantage of epitaxial growth to other methods is large-scale uniform production capability and more compatibility with current silicon processing techniques. Since SiC has been a semi-insulating substrate, the grown graphene layer on SiC can be used for device manufacturing [16, 17]. The characteristics of this epitaxial layer have been studied by different methods such as Atomic force microscopy (AFM)[17,18], Low-energy electron diffraction(LEED)[16], Raman spectroscopy[19-21], X-ray photoelectron spectroscopy (XPS), and scanning tunneling microscopy (STM)[22] Among these techniques, Raman spectroscopy is sensitive to the number of graphene layers and it is promising for an accurate thickness measurement[21,23]. This technique is an effective tool for understanding the behavior of phonons and electrons in graphene and provides valuable information about the degree of disorder and defect (D band), crystalline size along with the plan or in-plane vibration of $sp^2$ carbon atoms(G band) as well as the degree of stacking order(2D band)[20,24-27].

In this research, we aim to study the advantages and disadvantages of the physical vacuum deposition (PVD) method for growing the single layers graphene clusters theoretically and experimentally. In the theoretical study, we consider three types of graphene clusters and find their Raman spectrum and Infrared intensities versus wave number using the Hartree-Fock method. It is shown that the infrared intensity depends on the charge distribution on the cluster due to its geometry. The experimental infrared intensity shows some deviations from the ideal curve which can be attributed to one of the effective parameters which is the charge distribution on the cluster. The experimental Raman spectrum includes D-band, G-band, and 2D-band peaks in most graphene clusters but in some specific ones only G- and 2D-band are seen. An agreement was seen between the experimental and theoretical Raman spectrum when the wave number is less than 1700 $Cm^{-1}$. It is shown that for using the PVD method for growing the single-layer graphene clusters the uniformity of the temperature and the vacuum level of the chamber should be increased.

## 2. Theory and Calculation Method

### 2.1. Theory of Infrared Spectroscopy

According to the theory of Infrared (IR) Spectroscopy [28], when a molecule absorbs the infrared light with energy $h\nu$ its vibration energy changes according to the below equation

$$h\nu = E_n - E_m \tag{1}$$

Where, $E_m$ and $E_n$ are vibration energy. In this framework, the $x$-component of the electric dipole moment $\mu_x$ is given by [28]

$$(\mu_x)_{mn} = \int_{-\infty}^{\infty} \Psi_n \mu_x \Psi_m \, dQ \tag{2}$$

Where $\Psi$ denotes the eigenfunction of the molecule and Q denotes a normal coordinate. Therefore, IR absorption depends on the dipole moment and in consequence the charge distribution. It can be shown that [28]

$$(\mu_x)_{mn} = (\mu_x)_0 \int_{-\infty}^{\infty} \Psi_n \Psi_m \, dQ + (\frac{d\mu_x}{dQ})_0 \int_{-\infty}^{\infty} \Psi_n Q \Psi_m \, dQ \tag{3}$$

Where, the first-term denotes the magnitude of the permanent dipole of the molecule, and the second term is attributed to the absorption of the infrared radiation by the molecule.

### 2.2. Density Functional Theory (DFT)

For doing the theoretical study about the Raman spectra and the IR spectrum of single-layer graphene, we consider three types of clusters (see Fig.1) with the different amounts of molecules and different edge shapes and use the Density Functional Theory [29]. Of course, there are many types of clusters that can be chosen. But since our main purpose has been to show the geometry dependency of charge distribution and its effect on the IR intensity, we only consider these three types, as examples. Before calculating the IR intensities, the geometry of clusters should be optimized i.e., their ground state should be found. The most optimization algorithm implements the Hessian matrix which refers to the second derivative of the energy concerning the molecular coordinates. Gaussian Code considers the four convergence criteria during geometrical optimization [30]. The four convergence criteria are forces, the calculated displacement, and the root-mean-square (RMS) of both of them which must be essentially zero or very near to zero (i.e., 0.0045 for force, 0.003 for RMS of force, 0.0018 for displacement, and 0.0012 for RMS of displacement) [30]. Gaussian Code was used for optimizing three types of clusters (see Fig.1) and finding Raman spectra and infrared intensities. B3LYP functional and 3-21G basis functions were applied for finding the completely relaxed-structure of three types of clusters. After finding the relaxed structures, Raman spectra, and infrared intensities versus wave number were found by using the Freq Test Command of the Gaussian Code [30]. It should be noted that for finding them, the Hartree-Fock method was used. Using Gaussian View Code [30], one can find the highest occupied molecular orbital (HOMO), the lowest unoccupied molecular orbital (LUMO), Raman spectrum curve, and infrared intensities curve. It is noted that raw frequency values compared at the Hartree-Fock level contain known systematic error due to the neglect of electron correlation, resulting in an overestimation of about 10%-12% [30]. Therefore, it is usual to scale frequencies predicted at the Hartree-Fock level by an empirical factor 0.8929[30,31].

### 3. Experimental Details

The vacuum evaporation system was equipped with a heating source accessory for annealing the SiC wafers. The epi-ready and ultra-cleaned N-doped 6H-SiC (0001) Si-face wafer is placed in a tungsten boat and the boat is connected to the two electrodes of the heating source accessory. The SiC wafer is annealed based on the special recipe shown in table 1. Before annealing the SiC wafer, the chamber of the system is vacuumed by using rotary and diffusion pumps up to the $10^{-6}$Torr. The temperature of the boat is measured by using a thermocouple. It should be noted that, before heating the SiC wafers, the relation between the current passing between the two electrodes of the heating source accessory and temperature is found and the thermocouple is calibrated. Also, we use two types of laser during Raman spectroscopy. One laser with $\lambda$=532 nm and P=30 mW and another laser with $\lambda$=785 nm and P=50 mW.

## 4. Result and Discussion

Fig.2 shows the HOMO and LUMO of three types of clusters. As Figs. 2(b) and (c) show, these orbitals are localized at the up and down and left and right edges of the clusters shown in Fig.1 (B) and (C). Therefore, the charge distribution is not uniform on the cluster surfaces, and it is expected that the high infrared intensities will be found after doing the DFT calculations.

Fig.3 denotes the infrared intensities versus wave number of three types of clusters shown in Fig.1. As Fig.3 shows, the C-C and C=C bonds are seen in the range 1100-1700Cm$^{-1}$ and the C-H bond is seen near 3300 Cm$^{-1}$. As Fig.2 (a) shows the charge distribution is uniform on the surface of the cluster (Fig.1 (A)), approximately and its IR intensities are low (Fig.3 (a)) concerning the other two clusters. By comparison between Fig.2(c) and (b) it can be concluded that the charge distribution on the cluster shown in Fig.1 (B) is more uniform than the cluster shown in Fig.1(C). Therefore, as Fig.3(c) shows the IR intensities of Fig.1(C) is less than Fig.1 (B). Therefore, the charge distribution on the clusters can be considered as an effective factor for deviating the experimental results from the ideal ones when the IR spectrum is measured.

Fig.4 shows the Raman spectrum of the three clusters. The Raman spectrum depends on the type of coupling and vibration direction of neighborhood atoms concerning each other. Since we consider three small single-layer clusters, the coupling types and the vibration direction of atoms concerning each other specify the Raman spectrum. The $C - C, C = C$, and $C - H$ bonds are seen. It should be noted that the Raman spectrum of the clusters differs from the ideal single and multilayer graphene sheet due to the differences in numbers of atoms, edge shape, and various directions of atoms concerning each other. It means that the results of the Raman spectrum can be used for evaluation of the quality of the growth (e.g. crystal or polycrystal type of grown layer) and the number of layers in the multilayer graphene [20].

Fig.5 (a) and (b) illustrate the schematic of the first sample of SiC wafer after thermal treatment, (based on the data of table 1) and its experimental Raman spectrum versus wave number, respectively. The size of the sample is equal to $2.5 \times 3 \, Cm^2$.

It should be noted that Raman spectroscopy depends upon the inelastic scattering of photons, known as Raman scattering. The laser light interacts with molecular vibrations, phonons, or other excitations in the system, resulting in the energy of the laser photons being shifted up or down. In our studies, one laser with $\lambda$=532 nm and P=30 mW and another laser with $\lambda$=785 nm and P=50 mW have been used. The method is an analytical technique where scattered light is used to measure the vibrational energy modes of a sample. Raman spectroscopy can provide both chemical and structural information, as well as the identification of substances through their characteristic Raman 'fingerprint'. When light interacts with molecules in a gas, liquid, or solid, the vast majority of the photons are dispersed or scattered at the same energy as the incident photons. This is described as elastic scattering or Rayleigh scattering. A small number of these photons, approximately 1 photon in 10 million will scatter at a different frequency than the incident photon. This process is called inelastic scattering, or the Raman Effect. Raman spectros copy allows the user to collect the vibrational signature of a molecule, giving insight into how it is put together, as well as how it interacts with other molecules around it.

The spectrum is measured in the gray region of Fig.5 (a). As Fig.5 (b) shows, not only C=C and C$\cong$ C bonds but also the unexpected bonds: C=S, C-($NO_2$), S-H are seen. The presence of unexpected bonds in the Raman spectrum proves that the chamber is not clean and there are some contaminations inside the chamber. Also, since the spectrum deviates from the ideal spectrum it can be concluded that the growth layer is no single layer (or multilayers) graphene.

Now, a question can be asked: whether the single layer (or multilayers) graphene has been grown on the SiC substrate as a flake? Fig.6 (a) and (b) show the schematic of the second sample of SiC wafer after thermal treatment (based on the data of table 1) and its experimental Raman spectrum versus wave number, respectively. The second sample is prepared after cleaning the chamber. The size of the sample is approximately $2.5 \times 3 \, \text{Cm}^2$. The Raman spectrum on gray regions of the sample is almost similar to the first sample (it is not shown). But, let us consider the dark flake in the middle of the sample with dimension about $0.7 \times 0.8 \, \text{Cm}^2$, and get the Raman spectra of the flake [see Fig.6 (b)]. As Fig.6 (b) shows, not only the peaks related to the contaminations are omitted, but also the peak of G, D, and 2D, which are the main Raman characteristic of multilayer graphene[20] appear because the peak 2D is wide and its height is low[20].

Therefore, it can be concluded that the evaporation chamber should be completely clean, firstly. It can be done by cleaning the chamber with suitable detergents and reaching a higher vacuum level by using the molecular pump. Second, the temperature gradient on the SiC substrate and controlling the temperature increment and decrement during the thermal treatment should be more accurate for upcoming studies. Otherwise, the multilayer graphene will be only grown as a small flake on the SiC substrates (black flakes shown in Fig. 6(b)).

## 5.   Conclusions

The Gaussian Code was used and the Raman spectra and infrared intensities versus wave number of three types of well-relaxed graphene clusters (Fig.1) were found. It has been shown that when the non-uniformity of the charge distribution on the cluster surfaces is high (Fig.2 (b) and Fig.2(c)) the infrared intensity is high (Fig.3 (b) and Fig.3(c)). But, when the charge distribution is uniform, approximately (Fig.2 (a)), the IR intensity is low (Fig.3 (a)). Therefore, non-uniformity of charge distribution on the clusters is one of the effective factors which causes the differences between theoretical and ideal IR spectrum and experimental results. The theoretical Raman spectrum of these three different clusters has been found (Fig.4). Since, the number of atoms, edge shape, atomic coupling type, and the vibration direction of neighborhood atoms to each other have been different in each cluster concerning the other clusters, Figs.4(a), (b), and (c) are not similar to each other. It means that the Raman spectrum depends on the above parameters mentioned and it can be used to recognize the crystal or polycrystal type of grown layer and its number of layers in graphene case. Two samples of SiC wafer with dimension $2.5 \times 3 \, \text{Cm}^2$ (Fig. 5(a) and Fig.6 (a)) were annealed in a vacuum evaporation system equipped with heating source accessories based on the special thermal recipe (Table 1). It has shown that the contaminations were appeared as extra peaks in the spectrum (Fig.5 (b)). By removing the contamination, the extra peaks were deleted (Fig.6 (b)). Also, the Raman spectrum on the most area of the grown layer deviated from the ideal single or multilayer graphene (Fig.5 (b)), but on the small flakes, the spectrum coincided with the spectrum of multilayer graphene (Fig.6 (b)). The differences were attributed to eliminating the contaminations, the control of the temperature uniformity on

SiC substrate, and increasing and decreasing the temperature during the heat treatment. Therefore, by cleaning the chamber with suitable detergent, using the molecular pump for increasing the vacuum level inside the chamber, and using a more accurate temperature controller one can improve the quality of the grown layer.

**Figure caption**

**Fig. 1** (Color online) completely relaxed-structure of three types of graphene clusters

**Fig. 2**(Color online) HOMO and LUMO of three types of graphene clusters related to (a) Fig.1(C), (b) Fig.1(B), and (c) Fig.1(A).

**Fig.3** (Color online) Infrared intensities of three types of graphene clusters related to (a) Fig.1(A), (b) Fig.1(B), and (c) Fig.1 (C).

**Fig.4(Color online)** Theoretical Raman spectrum versus wave number of three types of graphene-like clusters shown in (a) Fig.1 (A), (b) Fig.1 (B), and (c) Fig.1 (C).

**Fig.5 (**Color online) (a) The schematic of the first sample of SiC wafer after thermal treatment and (b) the experimental Raman spectrum. The specification of the used laser in the Raman experiment is λ=532 nm and P=30 mW.

**Fig. 6 (Color online)** (a) The schematic of the second sample of SiC wafer after thermal treatment and (b) the experimental Raman spectrum. The specification of the used laser in the Raman experiment is λ=785 nm and P=50 mW.

6. **Table caption**

**Table 1- Temperature recipe for heating the SiC wafer in a vacuum chamber**

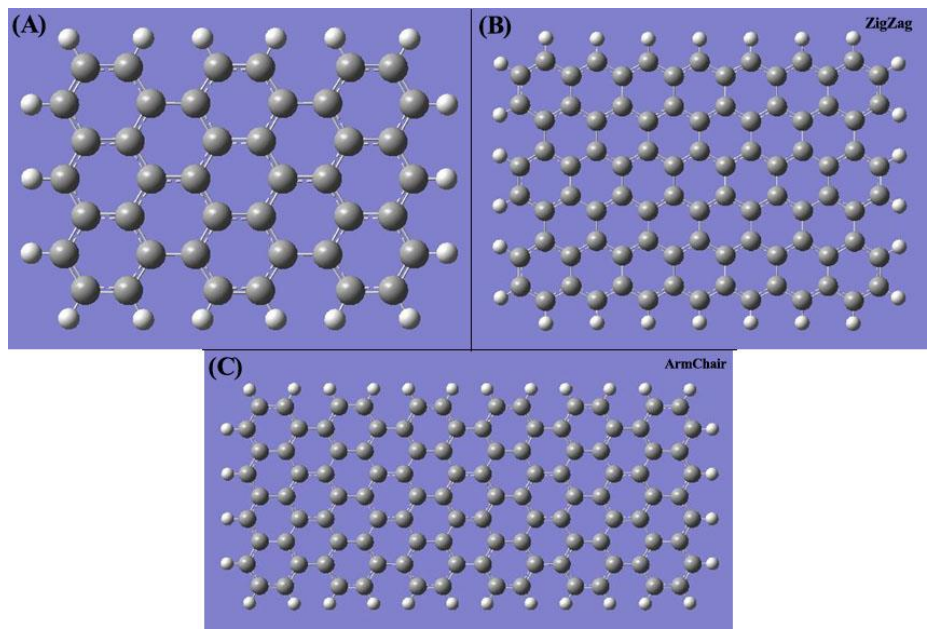

**Fig. 1 (Color online) completely relaxed-structure of three types of graphene clusters**

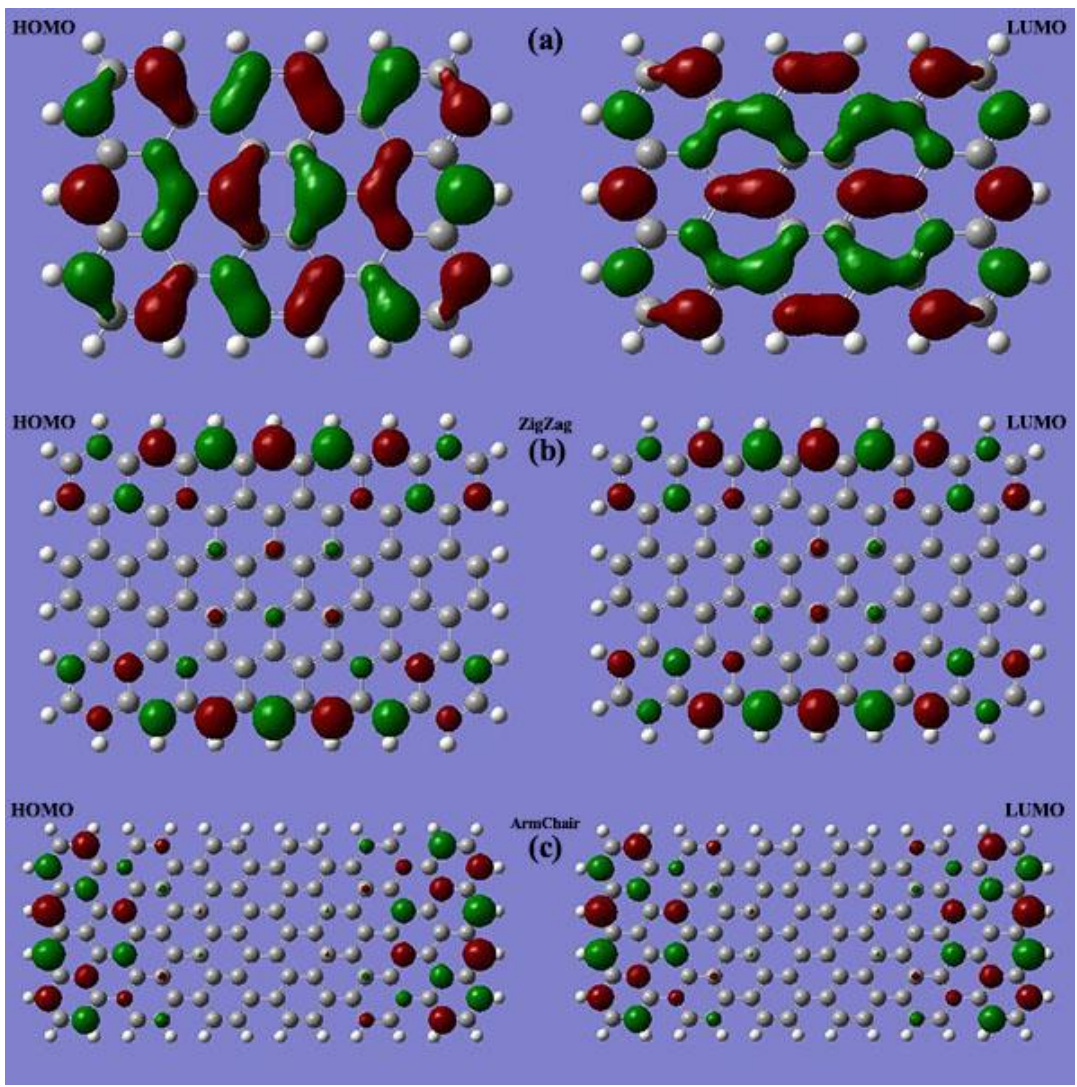

**Fig.2 (Color online) HOMO and LUMO of three types of graphene clusters related to (a) Fig.1(A), (b) Fig.1(B) and (c) Fig.1(C)**

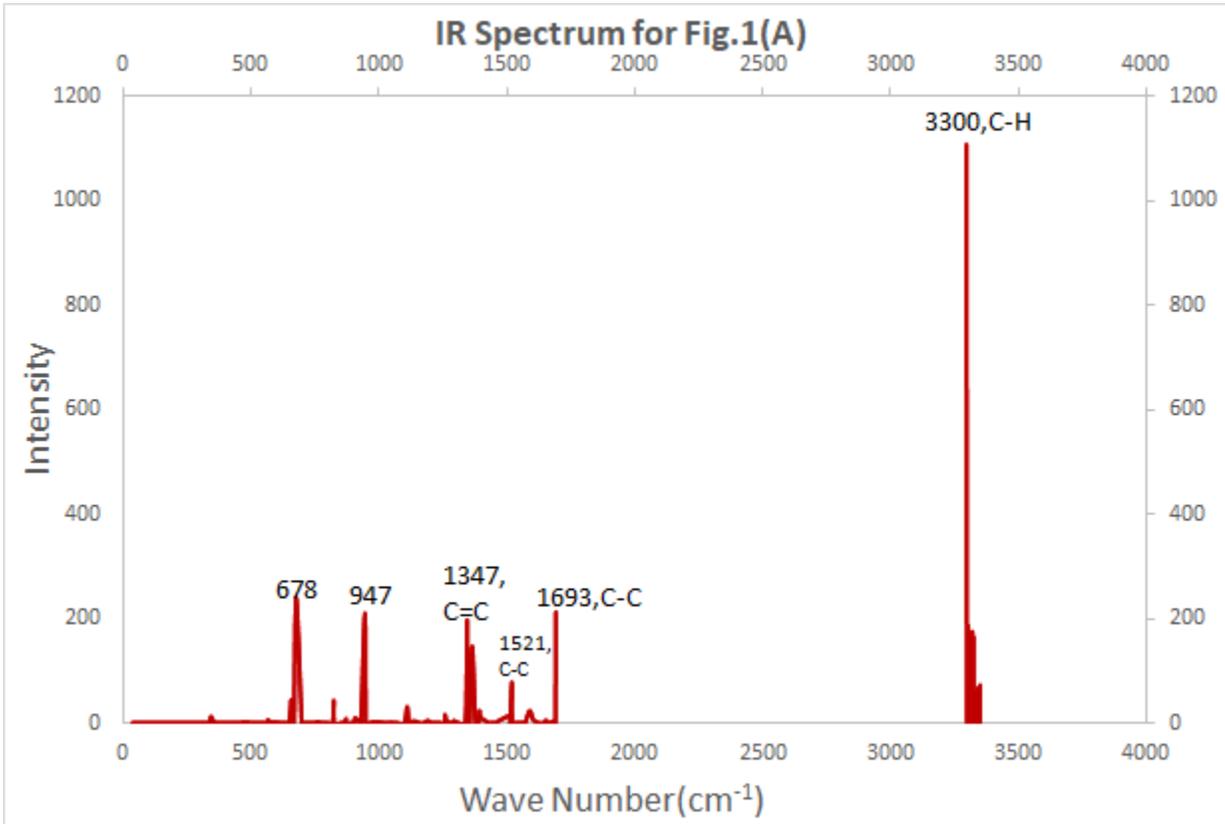

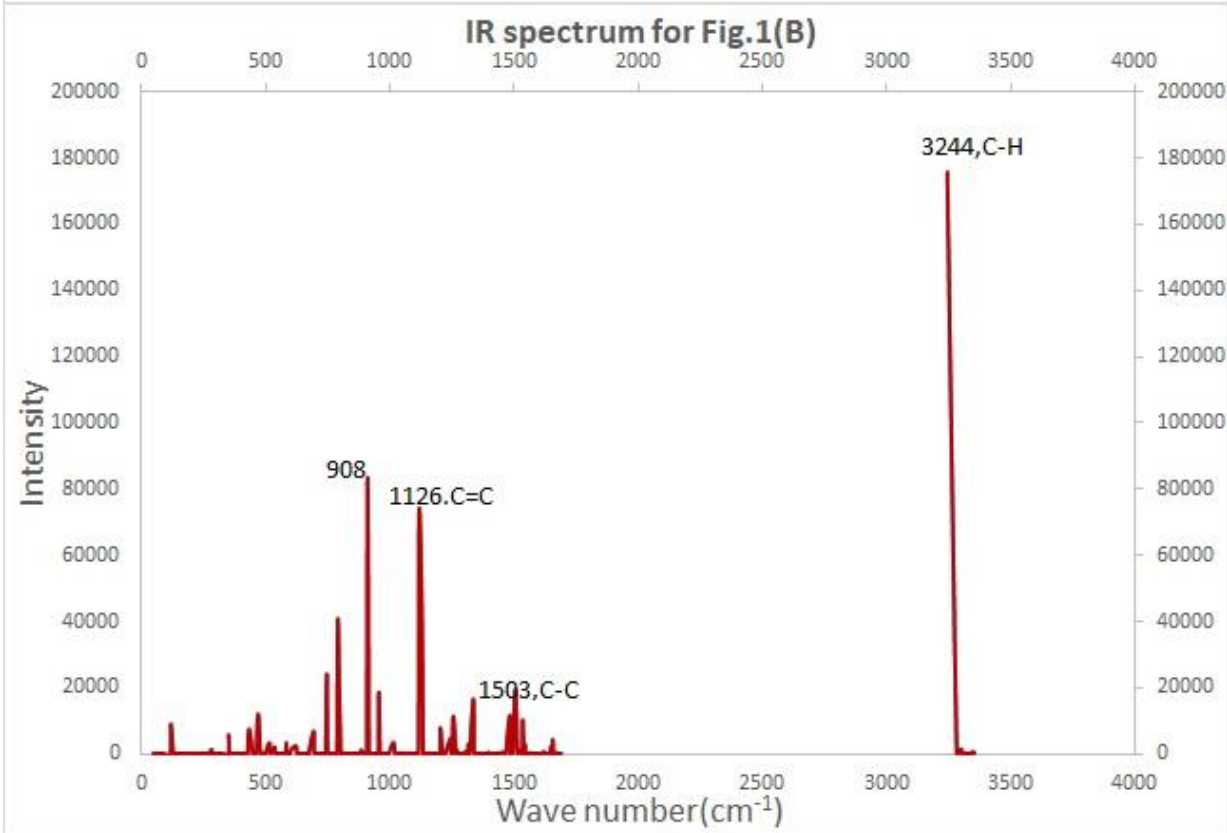

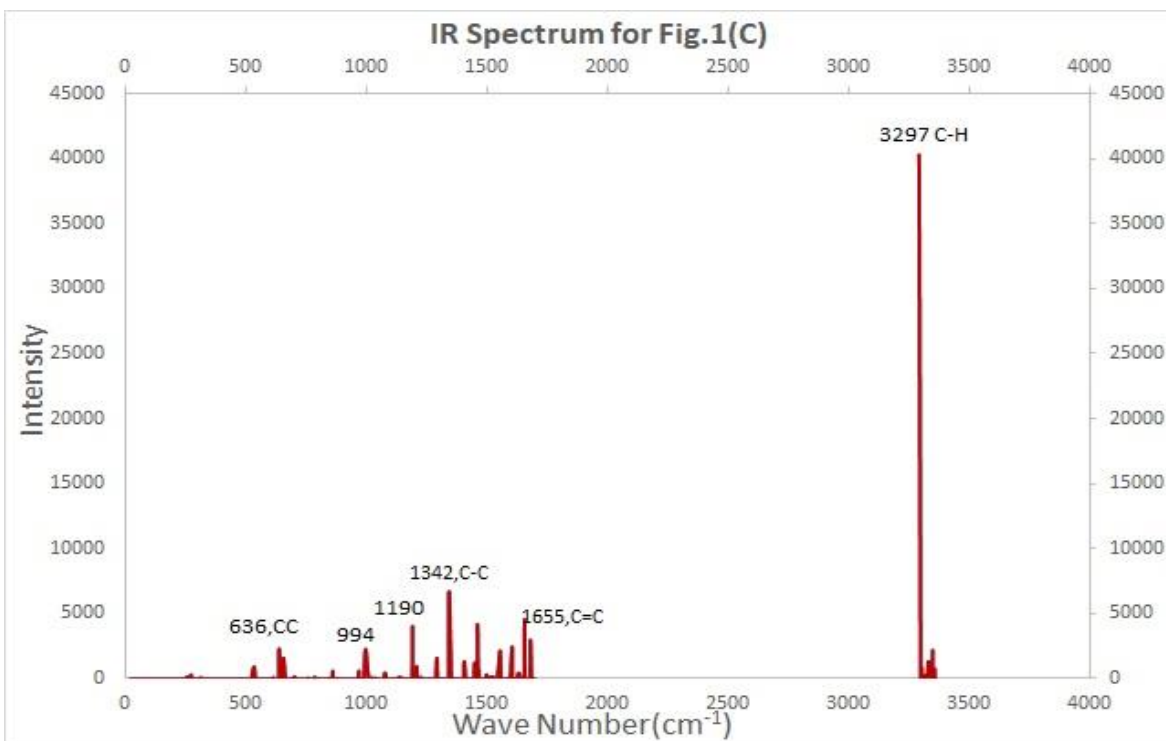

**Fig. 3 (Color online) Infrared intensities of three types of graphene clusters related to (a) Fig.1 (A), (b) Fig.1 (B), and (c) Fig.1 (C).**

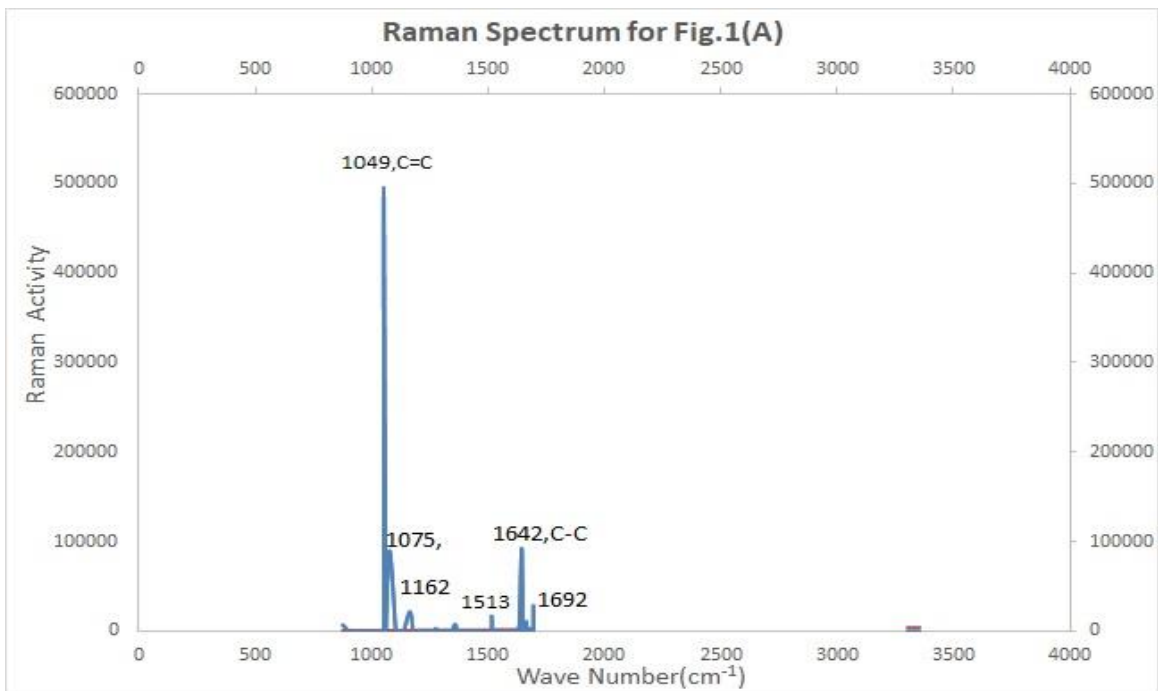

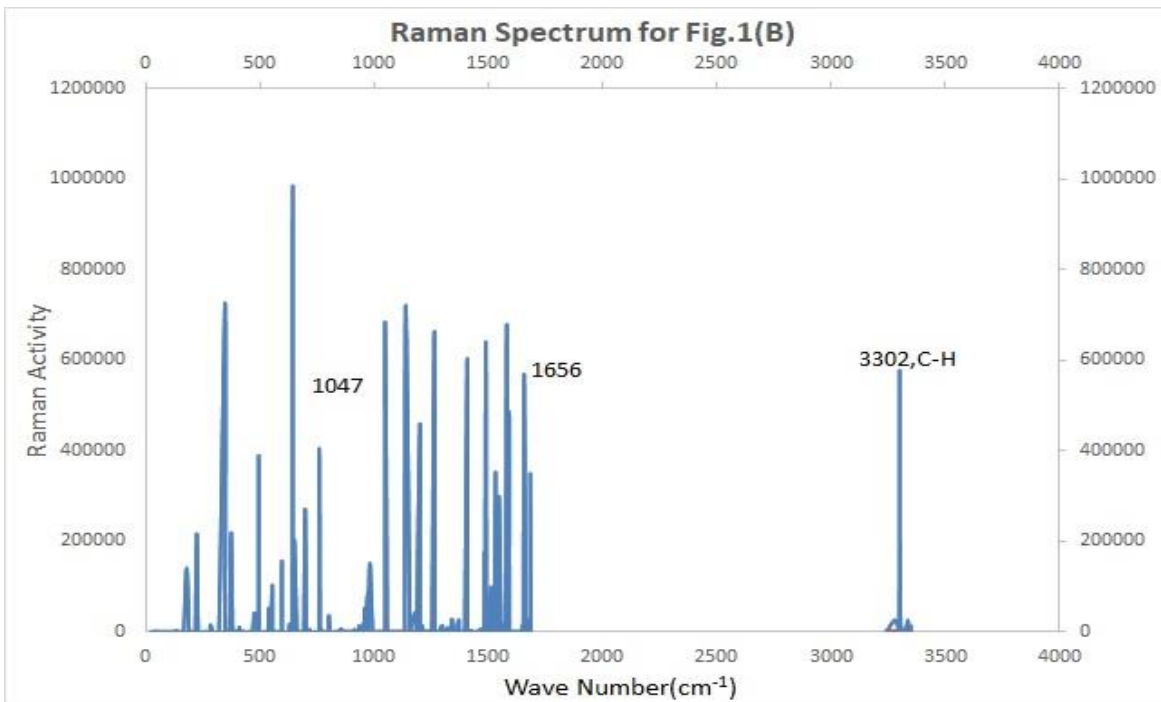

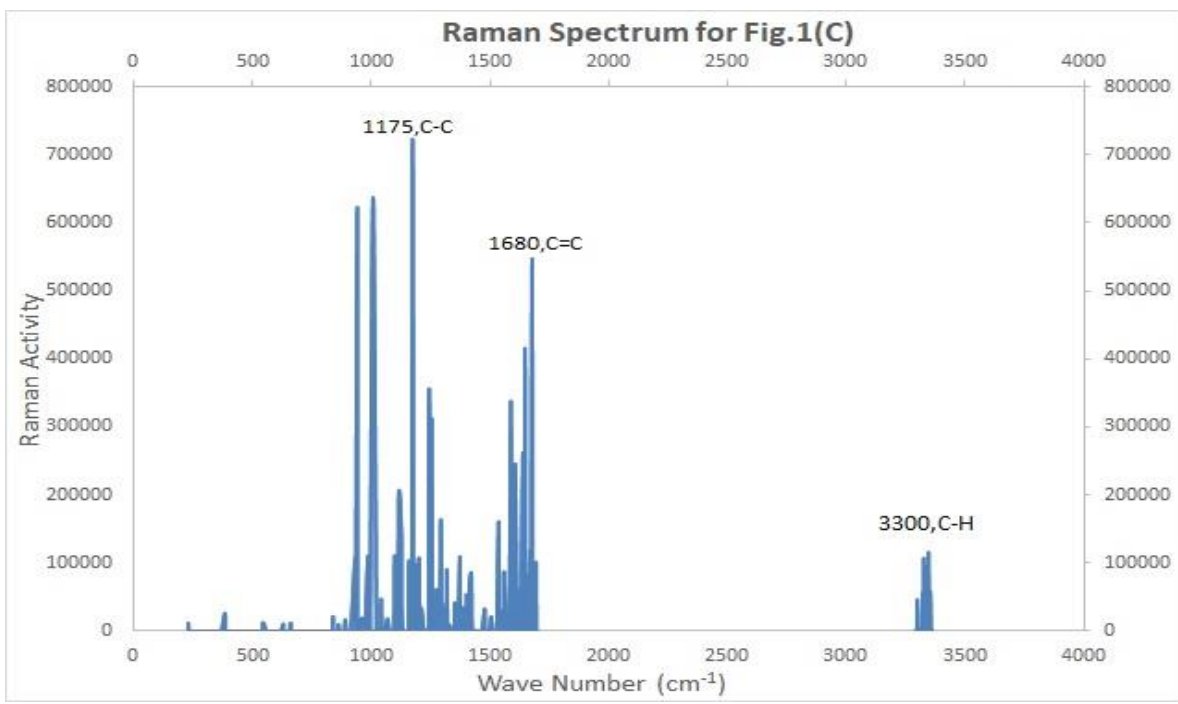

**Fig. 4 (Color online) Theoretical Raman spectrum versus wave number of three types of graphene-like clusters shown in (a) Fig.1 (A), (b) Fig.1 (B), and (c) Fig.1 (C).**

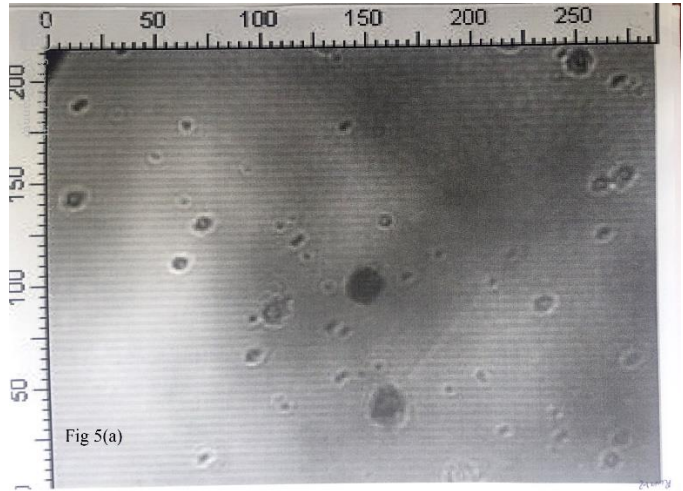

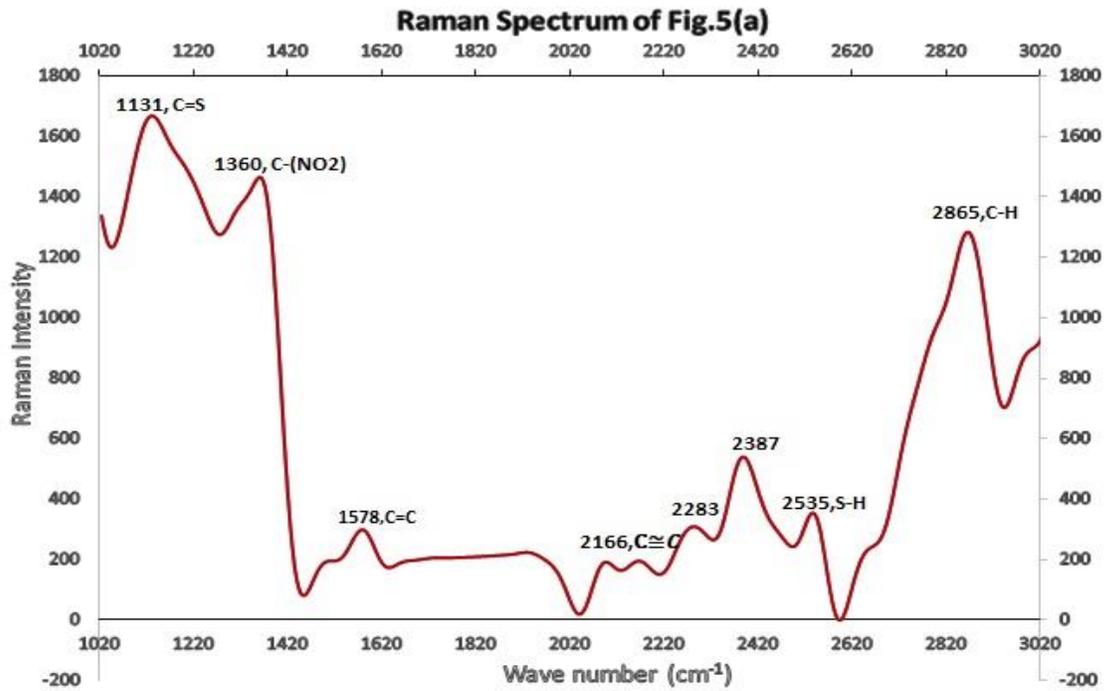

**Fig.5 (Color online) (a) The schematic of first sample of SiC wafer after thermal treatment and (b) the experimental Raman spectrum. The specification of used laser in Raman experiment is λ=532 nm and P=30 mW.**

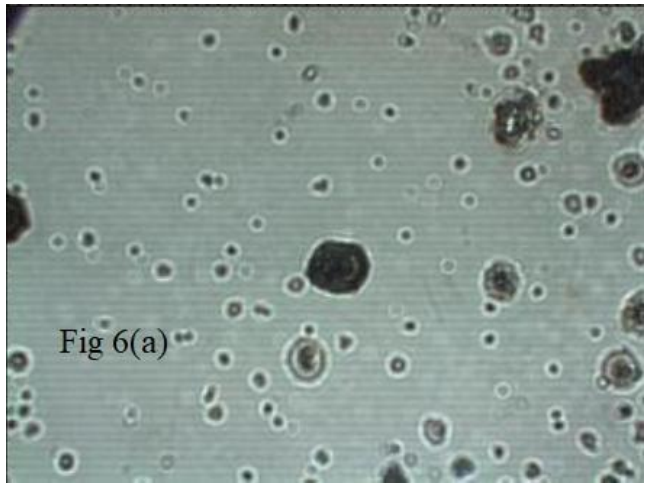

Fig 6(a)

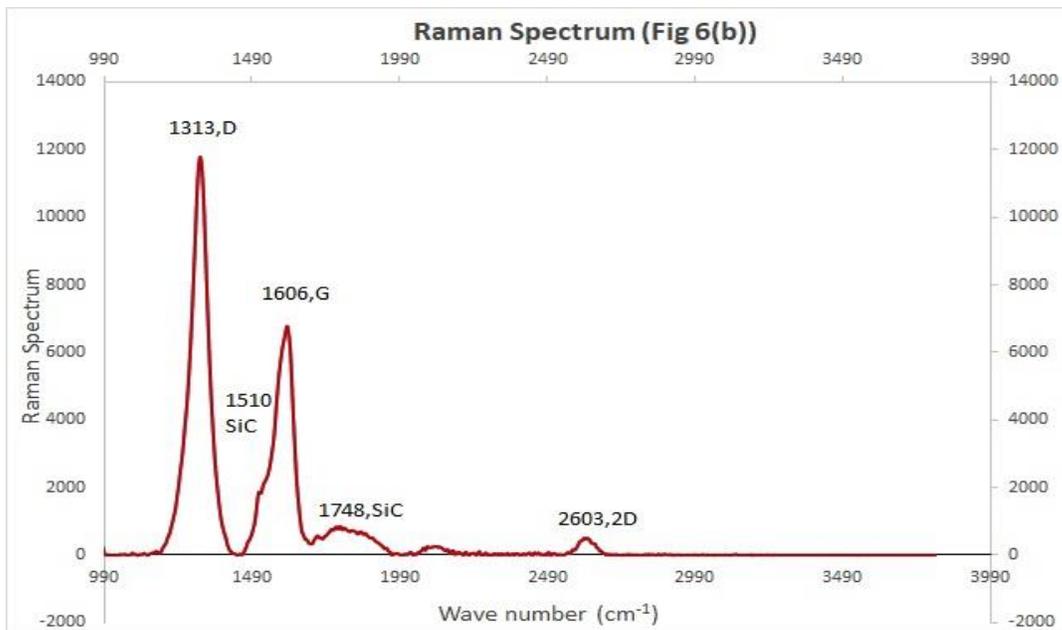

**Fig.6(Color online) (a) The schematic of second sample of SiC wafer after thermal treatment and (b) the experimental Raman spectrum. The specification of used laser in Raman experiment is λ=785 nm and P=50 mW.**

**Table 1- Temperature recipe for heating the SiC wafer in a vacuum chamber**

| Time (min) | Temperature($^0$C) |
|---|---|
| 0 | 41 |
| 1 | 800 |
| 2 | 1350 |
| 3 | 1438 |
| 7 | 1378 |
| 15 | 1258 |
| 20 | 1183 |
| 25 | 1108 |
| 30 | 1033 |
| 35 | 958 |
| 40 | 883 |
| 45 | 813 |
| 43 | 738 |
| 48 | 663 |
| 52 | 600 |
| 55 | 450 |
| 60 | 278 |
| 65 | 41 |